# Evidence of Metallic Wigner Crystal in Rhombohedral Graphene

Tonghang Han[1†], Jackson P. Butler[1†], Shenyong Ye[1†], Zhenqi Hua[2], Surajit Dutta[3], Zach Hadjri[1], Zhenghan Wu[1], Jixiang Yang[1], Junseok Seo[1], Phatthanon Pattanakanvijit[1], Emily Aitken[1], Kenji Watanabe[4], Takashi Taniguchi[5], Peng Xiong[2], Eli Zeldov[3], Zhengguang Lu[2], Raymond Ashoori[1], Long Ju[1*]

[1]Department of Physics, Massachusetts Institute of Technology, Cambridge, MA, 02139, USA.

[2]Department of Physics, Florida State University, Tallahassee, FL, USA

[3]Department of Condensed Matter Physics, Weizmann Institute of Science, Rehovot, Israel

[4]Research Center for Electronic and Optical Materials, National Institute for Materials Science, 1-1 Namiki, Tsukuba 305-0044, Japan

[5]Research Center for Materials Nanoarchitectonics, National Institute for Materials Science, 1-1 Namiki, Tsukuba 305-0044, Japan

*Corresponding author. Email: longju@mit.edu

†These authors contributed equally to this work.

**When the Coulomb interaction dominates over kinetic energy, electrons can crystallize into Wigner crystal (WC)[1]. Such a paradigmatic correlated electronic phase has been realized in two-dimensional electron gases (2DEGs) with a parabolic band dispersion[2–9] and completely flat Landau levels under high magnetic fields[10–14]. Beyond the conventional contexts of electron crystallization, more exotic electron crystals have been postulated but remain unexplored. For example, metallic Wigner crystal (mWC), a state with co-existing itinerant carriers and a pinned electron lattice, has been theoretically proposed but was thought to be challenging to realize[15–18]. Non-parabolic electron bands and quantum geometry could facilitate the realization of mWC and other novel topological electron crystals[19–28], offering new opportunities in an uncharted territory. Here we report transport evidence for WC and mWC in rhombohedral tetralayer, pentalayer, and hexalayer graphene in the charge density range of $0.3\text{-}0.5*10^{12}$ cm$^{-2}$. By flattening the conduction band with a gate-controlled displacement field *D*, we observe an insulating state at non-zero charge density that shows nonlinear and hysteretic current-voltage relation—signatures of a pinned WC[8,29] that are absent from the insulator at lower densities. Further increase of *D* from the WC state reveals transport properties dominated by hole-like carriers with only up to 15% of the nominal electron density, consistent with the expected phenomena of a mWC state. This mWC state is closely tied to the WC state, as they collapse simultaneously with increasing temperature or bias voltage. The mWC state exhibits a**

**quantum Hall effect onset of ~ 0.4 T and dis-obeys the Středa relation[30], indicating compressible exchange of charges between the itinerant holes and the transport-inert electron WC background. Our results establish rhombohedral graphene as a platform for exploring novel electron crystals, and possible non-trivial topology and new collective modes.**

Crystalline rhombohedral multilayer graphene (RMG) has emerged as a highly tunable system for emergent quantum phenomena. Two ingredients play central roles in making RMG a particularly fertile ground for correlated and topological electron physics. Firstly, the gate-controlled displacement field $D$ can continuously shape the low-energy bands, tuning not only the band dispersion and density of states but also the band's quantum geometric properties, including Berry curvature and orbital magnetic moment. This experimental knob enables the access to a band structure that combines flat and dispersive portions, a parameter space that is difficult to explore in other two-dimensional (2D) materials and conventional three-dimensional and two-dimensional electron gases (2DEGs). Secondly, electrons' internal spin and valley degrees of freedom enrich the flavor of ground states under strong electron interactions, resulting in a rich set of spontaneous isospin-symmetry-broken metallic[31,32] and insulating states[33–38]. Noticeably, when the conduction band bottom is tuned to be optimally flat and electrons populate part of this flat portion, orbital time-reversal-symmetry-breaking and associated chiral superconductivity have been reported[39,40]. At similar $D$ and $n_e$ conditions, additional moiré superlattice engineering between RMG and hexagonal boron nitride (hBN) induces fractional quantum anomalous Hall effects and fractional Chern insulators[41–45], hosted by the same electrons in the flat bottom of conduction band.

These same features of RMG also create new opportunities to explore unconventional crystalline electronic phases. Wigner crystal (WC)[1], a prototype strongly correlated phase with broken continuous translational symmetry, have been extensively studied in 2DEGs with either parabolic band dispersion at low densities[2–9] or with perfectly flat bands due to Landau quantization at high magnetic fields[10–14]. In contrast to the rigid band structures that host conventional WCs, electrical gating can tune RMG's band structure from a power-law dispersion with circular Fermi surface, to a flat bottom plus dispersive band, to a Mexican-hat-shaped dispersion with an annular Fermi surface (Fig. 1a), all featuring high electron density-of-states (Fig. 1b). Such tunable band dispersions, especially when the band bottom is optimally flat, could facilitate the formation of WC at zero magnetic field and provide a new testbed of the metallic (or self-doped) Wigner crystal, in which a pinned electron lattice coexists with a small density of mobile carriers[15–18]. Moreover, interaction-driven spin- and/or valley-symmetry-breaking provides a route to explore electron crystals with spontaneously broken time reversal symmetry. Theories have suggested that the tunable quantum geometry in RMG could facilitate electron-crystal phases with orbital magnetization and even non-trivial topology, such as chiral WC[46] and anomalous Hall crystal[19–28]. These exciting opportunities, however, have not been explored in experiments.

Here we investigate moiréless dual-gated RMG devices by magneto-electron transport measurements. Across tetralayer, pentalayer, and hexalayer RMG devices, we observed an insulating state that exhibits sharp threshold and pronounced hysteresis behavior with bias voltage at zero magnetic field. These features are both characteristic of a pinned WC. By increasing $D$ from this WC regime, the Hall response indicates that only up to 15% of the nominal (electron) charge density remains itinerant and behave as holes—phenomenologically consistent with a metallic (self-doped) WC state. In magnetic fields, the onset of Landau quantization and the evolution of the Landau fan further suggest a charge exchange between the itinerant hole-like carriers and an underlying crystalline electron reservoir. We have measured 2 tetralayer, 3 pentalayer and 1 hexalayer devices. In the main text we focus on the hexalayer device H1 and tetralayer device T1. Additional data, including pentalayer results, are provided in the Extended Data Figures.

**Gate-tuned WC in RMG**

Figure 1c shows the longitudinal resistance $R_{xx}$ as a function of electron density $n_e$ and displacement field $D$ at the base temperature 10 mK (See Methods). For $D/\varepsilon_0 \approx 0.65$–$0.82$ V/nm, a highly insulating regime is observed to span from zero $n_e$ to the lower-density-boundary of the chiral superconducting state[39]. The same measurement taken at 380 mK, however, reveals two separated regions inside this insulating state as shown in Fig. 1d: a lower-density region R1 connecting smoothly to the band insulator at zero $n_e$ and a higher-density region R2 centered at $(n_e, D/\varepsilon_0) \sim (0.4*10^{12}$ cm$^{-2}$, 0.79 V/nm). Figure 1e shows the continuous temperature evolution of $R_{xx}$ along the dashed lines in Fig. 1c&d, where R1 and R2 start to merge at ~300 mK. Although both reaching resistances above 1 MOhm at 10 mK, states in R1 and R2 show distinct transport properties through the current-voltage ($I$-$V_{bias}$) relation. Figure 1f shows the $I$-$V_{bias}$ curves taken at varied positions along the dashed line in Fig. 1c&d, corresponding to the colored ticks in Fig. 1e. At $n_e < 0.3*10^{12}$ cm$^{-2}$ (states in R1), $I$ exhibits nonlinearity versus $V_{bias}$, and the data for forward and backward scans of $V_{bias}$ overlap. In contrast, between $0.35*10^{12}$ cm$^{-2}$ and the onset of chiral superconductivity[39] (states in R2), $I$ shows a sudden turn-on at a threshold value of $V_{bias}$ to a state of linear $I$-$V_{bias}$, and a significant hysteresis is observed between the forward and backward scans of $V_{bias}$.

The contrasting $V_{bias}$-dependences in R1 and R2 point to distinct transport mechanisms and underlying ground states. R2 shows hallmarks of a WC state dominated by interaction-driven charge ordering: the sharp threshold and hysteresis are consistent with collective depinning and sliding of a disorder-pinned WC[8,29]. In contrast, states in R1 show behaviors of typical disorder-induced localization and variable-range hopping[47,48], and are continuously connected to the band insulator state at zero $n_e$. Although the crossover from disorder-induced localization to WC-like behaviors is not necessarily a sharp thermodynamic phase transition, the qualitative change of nonlinear response as a function of $n_e$ strongly supports identifying R2 as a pinned WC.

We calculated the effective $r_s$ as a function of $D$ and $n_e$ (see Methods), and found it can be more than 100 at the optimally flat band condition. This ratio between the Coulomb repulsion energy and averaged kinetic energy is sufficient to drive Wigner crystallization, supporting our observations of WC-like transport and assignment of states in R2 as WC. We note that the figures-of-merits observed in our experiment, e.g. the charge density corresponding to R1 and the melting temperature (~400 mK), are significantly higher than those of WC reported in other systems with similar transport signatures (ranging from $5*10^9$ cm$^{-2}$ to $1*10^{11}$ cm$^{-2}$, with melting temperatures below 100 mK)[3,7,8,49]. This can be understood by noticing the significantly larger effective mass $m^*$ (~3 times of bare electron mass, see Methods) and smaller dielectric constant κ (~4, of hBN) in our system. Since the Wigner-crystal melting temperature is expected to scale approximately as $m^*/\kappa^2$,[50] these material parameters can naturally lead to an enhanced melting temperature of WC in RMG.

We note that transport experiments might not differentiate the subtle scenarios in which the coherence length of WC is limited significantly by disorder effects, say the difference between Anderson Solid and Wigner Solid as suggested recently[51]. Although the coherence length of WC in graphene has been demonstrated to be at least hundreds of nanometers due to the low disorder density[13], it would still be valuable to obtain more direct evidence of electron crystallinity and coherence length using measurements other than DC transport. Beyond the sharp depinning reported here, signatures could come from narrow-band noise at microwave frequencies[52,53]; from tunneling spectroscopy[54]; from exciton Umklapp scattering by the electron crystal in rhombohedral graphene or a nearby transition-metal dichalcogenide sensing layer[5,6]; or from spatially resolved probes such as locally gated scanning tunneling microscopy[9,13] or nanoscale single-electron transistors[55].

Figure 1g shows the Hall resistance $R_{xy}$ as we scan the out-of-plane magnetic field at temperatures above the emergence of R2. In contrast to the insulating states below 400 mK, in which the $R_{xy}$ measurement suffers from the negligible current through the device, the device resistance at above 400 mK remains low enough to allow a reliable extraction of $R_{xy}$. Pronounced hysteresis and an anomalous Hall resistance are observed. The magnetic hysteresis and anomalous Hall effect observed in Fig. 1g suggest spontaneously broken time-reversal symmetry and development of valley polarization[31,39]. We will discuss the implications of this observed anomalous Hall effect on electron crystallizations later.

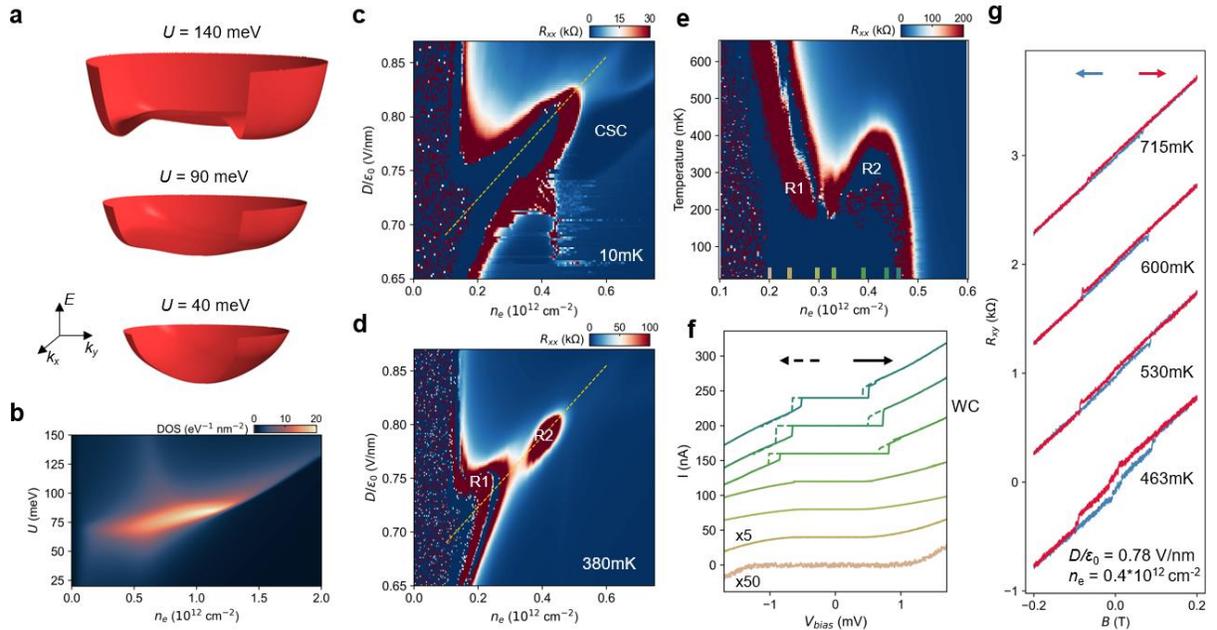

*Fig. 1 | Tunable Wigner crystal in rhombohedral graphene.* (a) Conduction-band dispersion of rhombohedral hexalayer graphene for interlayer potentials $U$ = 40, 90, and 140 meV, corresponding to the evolution from a monotonic E-k dispersion, to a flat disk with steep walls, to a Mexican-hat-shaped dispersion at the bottom of band. (b) Calculated Density-Of-States (DOS) versus electron density $n_e$ and interlayer potential $U$. (c) Longitudinal resistance $R_{xx}$ versus $n_e$ and displacement field $D$ at $B$ = 0 T and $T$ = 10 mK in device H1, showing a tilted, insulating region extending from the band insulator at zero density to the chiral superconductor (CSC) region at high-density. (d) Same map as in (c), but measured at 380 mK, revealing two disconnected regions (labeled as R1 and R2) within the insulating regime in (c). (e) $R_{xx}$ versus $n_e$ and temperature along the dashed line in (c) and (d). R1 and R2 are disconnected at higher temperature while they gradually merge below ~300 mK. (f) Current–voltage (I-$V_{bias}$) characteristics measured at states corresponding to the colored ticks in (e). In the higher density region R2, forward (solid) and backward (dashed) voltage scans show sharp turn-on and hysteresis, consistent with depinning of WC. In the lower density region R1, the current gradually turns on with increased bias voltage and no hysteresis is observed. (g) $R_{xy}$ measured during forward and backward scans of the out-of-plane magnetic field at temperatures higher than the emergence of the WC region R2, featuring anomalous Hall signals and valley-symmetry-breaking.

**Metallic (Self-doped) WC**

We next zoom into the neighborhood of the high-$D$ tip of the R2 region. Figure 2a&b show the $R_{xx}$ and $R_{xy}$ maps taken at an out-of-plane magnetic field $B = 0.1$ T. Strikingly, we observed a region (labeled as 'R3') with unusual Hall signal that is located between the WC region and the neighboring metallic states: the sign of $R_{xy}$ in this region is opposite from the sign of $R_{xy}$ in neighboring states. This Hall signal might be dominated by an anomalous Hall contribution, which can have either sign. To eliminate the anomalous signal, we measured $R_{xy}(B)$ with forward and backward field scans at four representative points (Fig. 2c–f). Aside from the hysteretic components, the normal Hall components of $R_{xy}$ in Fig. 2c–e are all electron-like and consistent with the type of nominal charge density by gating. In contrast, Fig. 2f shows a dominant hole-like normal Hall slope with a magnitude that is nearly an order of magnitude larger than that in Fig. 2d, despite similar nominal $n_e$. To map this behavior across the parameter space, we define a Hall density $n_{Hall}$ based on the difference of $R_{xy}$ between 0.2 T and 0.1 T, which eliminates the hysteretic/anomalous components (Fig. 2g). In the region R3, the extracted carriers are hole-like, with density increasing smoothly from zero at the WC boundary to ~0.06 × $10^{12}$ cm$^{-2}$ (~15% of the nominal charge density) as $D$ increases, before crossing over to a conventional metallic state at larger $D$. Figure 2h shows representative line-cuts at constant $n_e$ and $D$ corresponding to the dashed lines in Fig. 2g. Along the constant-$D$ cut, the hole density is changed by ~2*$10^{10}$ cm$^{-2}$ while the nominal electron density is changed by ~5*$10^{10}$ cm$^{-2}$—meaning slightly less than half of the added charges deplete the holes, and the rest becomes transport-inert electrons. We note that $n_{Hall}$ diverges near the phase boundary because $R_{xy}$ crosses zero as the charge transport transitions from being dominated by holes to by electrons.

This observation is counterintuitive at first glance: the nominal electron density exceeds the extracted hole density by an order of magnitude, yet the Hall response is hole-like. To further characterize the carriers contributing to transport in this regime, we analyze the low-field magneto-transport using a two-carrier Drude model (see Methods for details). By fitting $R_{xx}(B)$ and $R_{xy}(B)$ at the same time, while constraining the total charge density as the nominal gate-defined $n_e$, we can extract the electron and hole mobilities as shown in Fig. 2i. In the region R3, the electron mobility drops sharply, while the hole component exhibits an anomalously large mobility—not only compared to the electron mobility in the same region, but also compared to that of electrons in the adjacent metallic states.

These observations establish a phenomenon in which low-density itinerant holes co-exist with localized electrons with a higher density. When the $D$ changes, the electron sector serves as a charge-reservoir to exchange charges with the hole sector. When the total charge density $n_e$ is changed, some electrons are added to or removed from the localized electron reservoir, and the rest of the change of $n_e$ is absorbed into the itinerant hole sector. Since the region R3 resides right next to the R2 region of WC, we interpret the states in R3 as metallic Wigner crystals (mWC): electrons are localized into a WC that is pinned by disorder, therefore not carrying current or contributing to the Hall signal, but serving as a charge reservoir; at the same time, co-existing hole-like carriers move in the background of electron WC, support metallic charge transport, and exchange charges with the transport-inert reservoir of electron WC.

A mechanism to generate such mWC has been proposed theoretically[18]: quantum dynamics of point defects (vacancies and interstitials) in WC can drive a self-doping instability and stabilize an intermediate "metallic electron crystal" near the WC–metal transition (Fig. 2j). Parabolic bands, however, were thought to be hard to accommodate such mWC—possibly explaining its absence from experiments so far[15,16]. In rhombohedral graphene, the single-particle band structure may favor the self-doping mechanism: as $D$ is increased continuously, the Fermi surface (in the conduction band) evolves from a circular shape, to an almost perfectly flat disk with steep walls[56], to a Mexican-hat shape. The almost perfectly quenched kinetic energy of the flat bottom of the conduction band provides a reasonable setting of electron crystallization and might enable the self-doping mechanism. Phenomenologically, the region R3 resides at slightly higher $D$ values than that of R2, which might suggest some role played by the emergent annular Fermi surface in the Mexican-hat-shaped band structure in facilitating mWC in region R3. However, we note that this simple band-dispersion picture corresponds to the weak-coupling-limit of interacting electrons, which is different for the WC scenario corresponding to the strong-coupling-limit. Further theoretical analysis and numerical calculations are needed to investigate the unconventional flat disk plus steep wall dispersion and the Mexican-hat-shaped dispersion, both of which are a new realm for electron crystallization research.

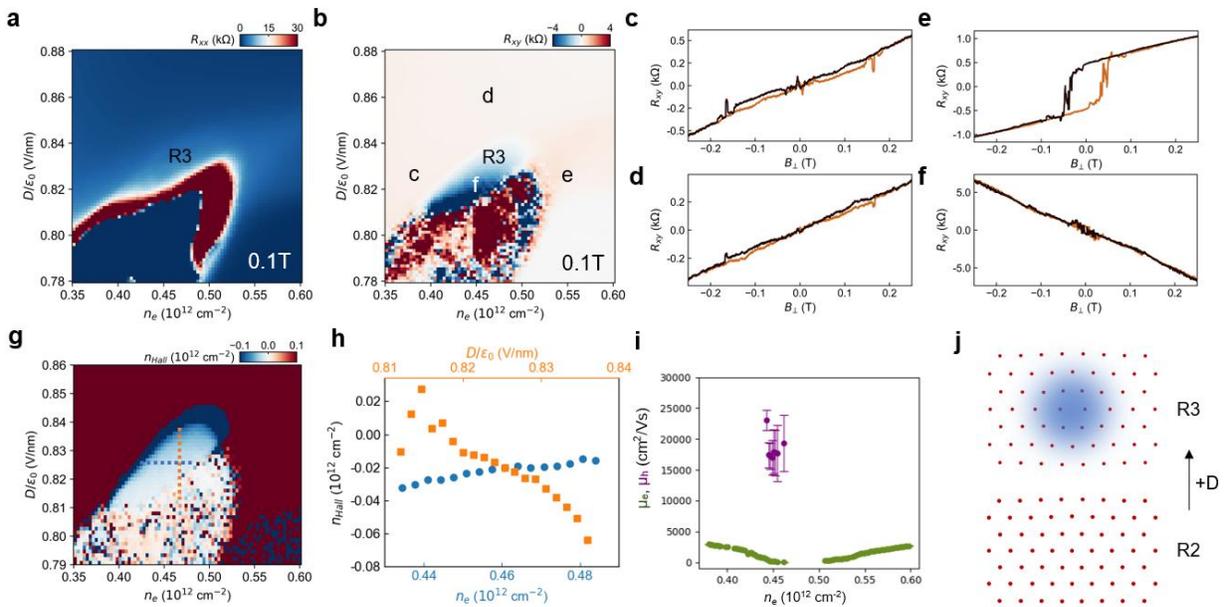

**Fig. 2 | Metallic Wigner crystal.** (a,b) Longitudinal ($R_{xx}$) and Hall ($R_{xy}$) resistances versus $n_e$ and $D$, measured at $T = 10$ mK and $B = 0.1$ T in H1. On the higher-$D$ side of the WC regime, a region (labeled as 'R3') with opposite Hall resistance from neighboring states is observed. (c–f) $R_{xy}$ versus magnetic field $B$ measured with forward and backward field scans at the locations marked in (b): ($n_e$, $D/\varepsilon_0$) = (0.38, 0.826) (c), (0.47, 0.86) (d), (0.55, 0.826) (e), and (0.45, 0.82) (f), in units of ($10^{12}$ cm$^{-2}$, V/nm). (g) Hall density $n_{Hall}$ extracted from the anti-symmetrized $R_{xy}$ using the difference between 0.2 T and 0.1 T. (h) $n_{Hall}$ along the dashed lines in (g) at $n_e = 0.467*10^{12}$ cm$^{-2}$ (bottom) and $D/\varepsilon_0 = 0.825$ V/nm (top). (i) Mobilities of electrons and holes at $D/\varepsilon_0 = 0.825$ V/nm extracted from a two-carrier model (see Methods). The electron-mobility in the region represented by (f) is close to zero, while the hole-mobility is ~ two orders of magnitude higher. Outside that region, the hole mobility cannot be reliably determined due to the much smaller density of holes tha electrons. These observations agree with the expected phenomena in a metallic WC with self-doped vacancies. (j) Schematics of the WC and metallic WC. Lower panel: WC with a triangular lattice of electrons (red dots), which corresponds to when electrons fill the flat bottom of the conduction band. Upper panel: mWC with a vacancy (blue blob) that is de-localized and carries positive charge, co-existing with localized electrons.

**Connection between mWC and WC**

We next examine the temperature and bias dependences of the mWC and WC. Figures 3a-f show $R_{xx}$ and $R_{xy}$ measured at $B = 0.2$ T at 380 mK (a,d), 470 mK (b,e), and 660 mK (c,f), respectively. With increased temperatures, the finite-density resistive region R2 (corresponding to WC) and the negative-$R_{xy}$ region R3 (corresponding to mWC) shrink and eventually disappear simultaneously, while the states in region R1 near zero charge density are much less sensitive to temperature changes. Figure 3h,i show the current $I$ and Hall voltage $V_{xy}$ as a function of DC bias voltage $V_{bias}$ at states corresponding to the colored dots in Fig. 3g, with two states residing in the WC region, two in the mWC region, and two in the Fermi liquid region. The current $I$ shows the typical turn-on behavior and sharp kinks at threshold bias voltages in the WC region, while it increases linearly in the Fermi liquid region at high $D$. In region R3, we observed similar kinks and nonlinear $I$-$V_{bias}$ relation as in the WC region. The Hall voltage $V_{xy}$ shows a monotonic increase with $V_{bias}$ in both the de-pinned WC and Fermi liquid regions, corresponding to electron-dominated transport. In contrast, a non-monotonic dependence of $V_{xy}$ on $V_{bias}$ is observed in the mWC region. The $V_{xy}$ changes sign at around the threshold voltage (corresponding to the kinks in Fig. 3h), indicating a transition from hole-dominated to electron-dominated transport before and after the depinning of WC.

The similar temperature and bias voltage dependences of transport in WC and mWC agree with the picture that they share a common electron crystal component. The temperature evolution indicates that the region R3 tracks the finite-density resistive region (R2) rather than the insulator close to zero charge density, supporting the interpretation that the region R3 is a self-doped continuation of the WC (R2). As for the bias voltage dependence, even with the current carried by effective holes in the mWC state running in the background, the depinning behavior of the coexisting WC of electrons can still be observed in experiment. These observations further strengthen the assignment of the hole-like transport region R3 as a mWC.

For comparison, we would like to also discuss a more trivial picture: electron-like and hole-like charge carriers originate from the inner and outer circles of the annular Fermi surface, when the Fermi level cuts through the Mexican-hat-shaped band bottom; without interactions, they remain as independent sectors that contribute to transport through a simple summation. This is different from what happens in the mWC state, where the hole-like charge carriers originate from self-doped vacancies in the electron WC. Firstly, we determined the effective mass of the hole-like carriers to be much smaller than that derived from the Mexican-hat-shaped band structure at both the inner circle and outer circle of the annular Fermi surface (see Methods). This is consistent with the observed exceptionally larger mobility of the hole-like carriers than that of electrons both inside the region R3 and in adjacent metallic states. Secondly, we calculated the expected Hall voltage for a bias voltage larger than the threshold value shown in Fig. 3h, using a two-carrier model (see Methods for details). By choosing realistic values of electron and hole mobilities, we found that the Hall voltage should still be dominated by holes at beyond the depinning threshold, which is incompatible with the observed sign change of $R_{xy}$ at threshold bias voltages. Thirdly, at temperatures above the disappearance of R2, we expect the independent electron-like and hole-like Fermi surfaces to contribute to transport through a simple summation. However, the observed Hall resistance is dominated by electrons in the region R3 as shown in Fig. 3f, same as the observation at base temperature with bias voltages beyond the depinning threshold. All these findings clearly suggest that the hole-like carriers interact strongly with the disorder-pinned electron crystal, and their mass and mobility are only defined when the mWC is formed, or at least significantly altered from values derived from a simple non-interacting band picture.

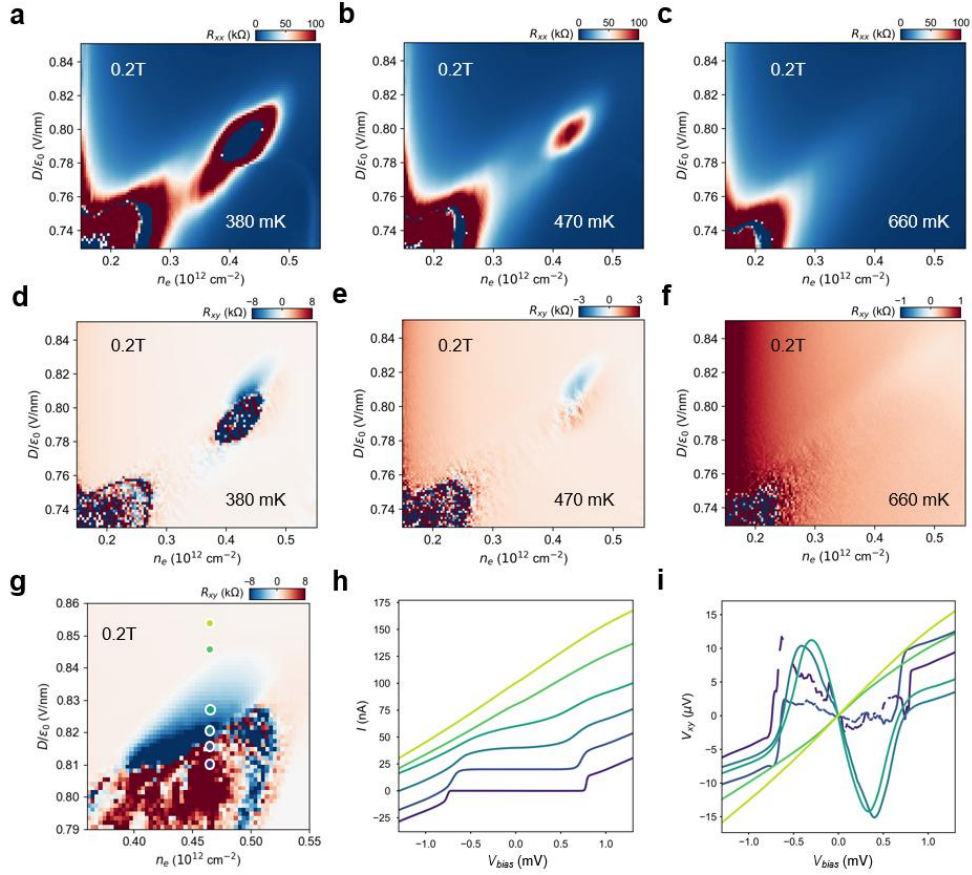

*Fig. 3 | Temperature and bias voltage dependences of WC and mWC in device H1.* (a,d) $R_{xx}$ and $R_{xy}$ versus $n_e$ and D, measured at T = 380 mK and B = 0.2 T. (b,e) same as (a,d) but measured at T = 470 mK. (c,f) same as (a,d) but measured at T = 660 mK. The WC and mWC regions emerge and strengthen together as the temperature is cooled down, suggesting their closely connected microscopic origins. (g) $R_{xy}$ versus $n_e$ and $D/\varepsilon_0$, measured at T = 10 mK and B = 0.2 T. (h,i) Current I and Hall voltage $V_{xy}$ as a function of DC bias voltage $V_{bias}$, taken at six states as labeled by colored dots in (g). At the two states inside the WC region R2, clear turning-on of current at threshold bias voltage and fluctuating Hall voltage below the threshold are observed. At the two states in the Fermi liquid region, simple linear I-$V_{bias}$ relation and monotonic electron-dominated Hall voltage are observed. At the two states in the region R3, 'kink' in the I-$V_{bias}$ relation at threshold bias voltage similar to that in the WC region is observed. Simultaneously, a non-monotonic $V_{xy}$ that changes sign at about the same threshold bias voltage is observed. These observations suggest coexisting electron WC and itinerant holes, agreeing with the mWC by self-doping picture.

**Quantum Hall effect of the mWC**

Figures 4a&b show $R_{xx}$ and $R_{xy}$ data around the mWC region from device T1, at out-of-plane magnetic field $B$ = 1 T. The mWC region splits into two regions with $R_{xy}$ quantized to $-h/e^2$ and $-h/(2e^2)$, corresponding to Chern number $C$ = -1 and $C$ = -2, respectively. Figures 4c&d show $R_{xx}$ and $R_{xy}$ versus $n_e$ and $B$ at fixed $D/\varepsilon_0$ = 1.043 V/nm, and Figs. 4e,f show the corresponding data versus $D$ and $B$ at fixed $n_e$ = 0.34*10$^{12}$ cm$^{-2}$. In both Landau fans, a cascade of quantized $R_{xy}$ states with $C$ = -1, -2, and -3 develops as $B$ is increased (see Extended Data Fig. 3 for more details). Beyond a magnetic field of up to 2.5 T, the sign of $R_{xy}$ changes from hole-like to electron-like. Similar sign changes of $R_{xy}$ are observed by applying a DC depinning bias voltage or by warming up the sample under non-zero magnetic field (see Extended Data Fig. 3). Figure 4g&h show temperature dependence of $R_{xx}$ and $R_{xy}$ taken at $n_e$ = 0.34 × 10$^{12}$ cm$^{-2}$ and $B$ = 1 T. At low temperatures, $R_{xy}$ is quantized at $-h/e^2$ and is almost quantized at $-h/(2e^2)$, corresponding to $C$ = -1 and -2, respectively. At high temperatures, such quantization disappears and the sign of $R_{xy}$ is changed to electron-like.

      The quantized Hall resistance and corresponding vanishing longitudinal resistance indicate a clear quantum Hall phenomenon due to Landau quantization of the motion of holes in the mWC state. The details, however, are unusual in three ways. Firstly, the evolution of region with quantized Hall resistance does not follow the Středa relation (indicated by the dashed lines). Secondly, the $n_e$–$B$ map lacks a conventional Landau fan: transitions between different Chern numbers occur at nearly the same magnetic field over a wide range of nominal $n_e$. In contrast, the $D$–$B$ map shows a fan-like structure, consistent with the density of itinerant (hole-like) carriers being controlled primarily by $D$ rather than by the total charge density. Thirdly, the density span of quantized plateaus is anomalously large: at 1 T the usual single-flavor Landau-level density spacing should be ~0.024 × 10$^{12}$ cm$^{-2}$, yet the observed plateau widths reach ~0.1 × 10$^{12}$ cm$^{-2}$. These features follow naturally from the charge-reservoir picture: a pinned electron crystal can continuously absorb or release carriers, so the phase boundaries are not only set by the filling of Landau levels, but also by the competition with the adjacent metallic states. The onset field for the quantum Hall effect in the MWC is also unusually low: the approximately 0.4 T onset is substantially below that of adjacent metallic states. Such a low onset implies a high mobility and a much smaller effective mass of the itinerant holes compared to calculations based on a tight-binding model (Extended Fig. 4). This observed small effective mass of vacancies in the mWC aligns with the predications by previous theory[18].

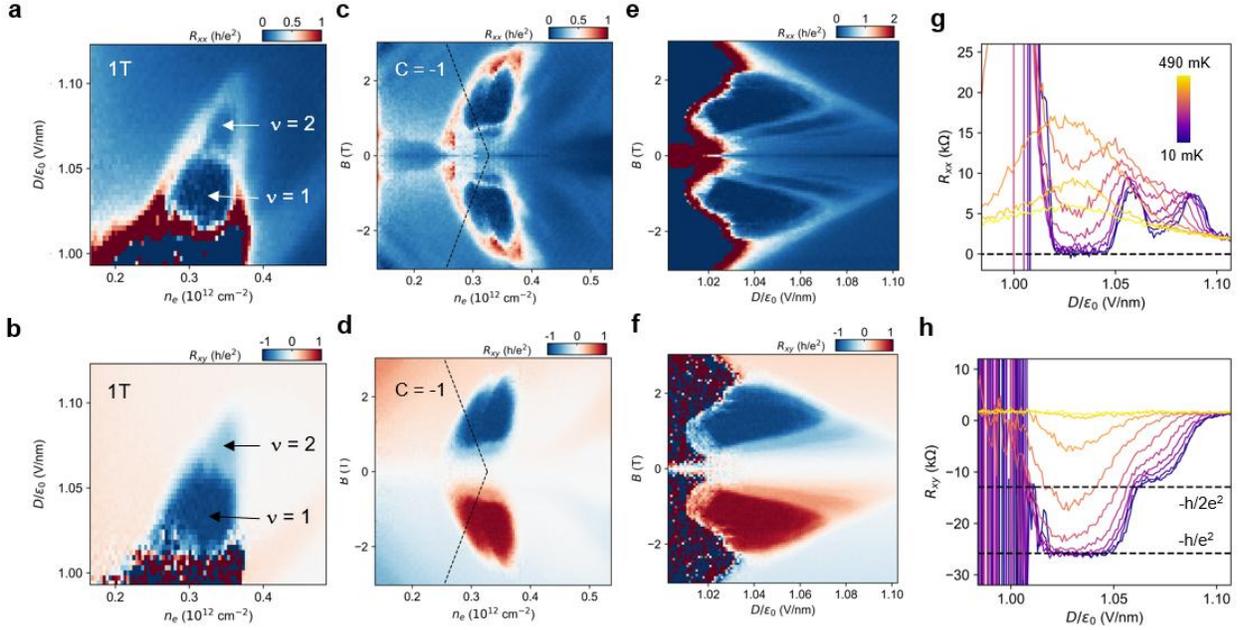

**Fig. 4 | Unconventional quantum Hall effect of the mWC.** (a,b) $R_{xx}$ and $R_{xy}$ versus $n_e$ and $D$ at $T = 10$ mK and $B = 1$ T in rhombohedral tetralayer graphene device T1. Two separate regions within mWC show uniformly distributed $R_{xy}$ of roughly $-h/e^2$ and $-h/(2e^2)$, respectively, and they both show vanishing $R_{xx}$. These observations indicate the Landau quantization of holes when 1 and 2 Landau levels are filled, respectively, while the electrons remain localized and inert in transport. The $v = 1$ region resides between the WC and $v = 2$ regions along the D-axis, consistent with the gradual increase of hole-density as a function of $D$ shown in Fig. 2g&h. (c,d) $R_{xx}$ and $R_{xy}$ versus $n_e$ and $B$ at fixed $D/\varepsilon_0 = 1.043$ V/nm. At low $B$ fields, the evolution of the Landau fan does not follow the slope defined by the Středa formula (black dashed lines, corresponding to $C = 1$ and 2, respectively). Above ~2 T, the $R_{xy}$ signal corresponding to holes disappears and the whole density range is dominated by electron-like $R_{xy}$ signals—indicating the collapse of mWC. (e,f) $R_{xx}$ and $R_{xy}$ versus $D/\varepsilon_0$ and $B$ at fixed $n_e = 0.335 \times 10^{12}$ cm$^{-2}$. Quantized $R_{xy}$ onsets at $B \approx 0.5$ T. (g,h) Temperature dependence of $R_{xx}$ and $R_{xy}$, taken at $n_e = 0.34 \times 10^{12}$ cm$^{-2}$ and $B = 1$ T. $R_{xy}$ is quantized at $-h/e^2$ and is almost quantized at $-h/(2e^2)$ only at low temperatures, while it changes sign at high temperatures. This agrees with the picture of holes dominating the transport in mWC while electrons dominating the transport when mWC melts with increased temperature.

## Discussion

Our observations show evidence of the electrically tunable band structure of rhombohedral graphene as an unconventional setting for novel electron crystals. One immediate intriguing question to ask is whether the valley polarization and time-reversal-symmetry-breaking exists in the WC itself. In conventional WCs from a parabolic band, the vanishing wave-function overlap suppresses exchange interaction and thus favors flavor-unpolarized order[57]; in RMG, the scenario might be different, as the WC develops from a high-temperature valley-polarized electron liquid state. We also note that theory suggests the possibility of spontaneous valley polarization of WC in multi-valley materials with trigonal warped Fermi-surface[58], even in the absence of spontaneously valley-polarized Fermi liquid to begin with[57]. These features of the rhombohedral graphene raise the prospect of "chiral" electron crystals with orbital magnetization[46]. Determining the magnetism and valley polarization of the WC in its ground state will require additional probes, for example, SQUID magnetometry[59] or optical probes such as Kerr rotation / RMCD[60]. But these techniques and experiments are beyond the scope of this work.

Furthermore, the possibility of integrating electron crystal with topology has been proposed by theories in the setting of rhombohedral graphene, as the anomalous Hall crystal (AHC)[19–28]. Such a topological electron crystal features chiral edge modes that support quantized anomalous Hall conductance. We would like to point out the difference between the observed mWC and an incipient AHC that needs additional assistance by magnetic field to show fully quantized anomalous Hall resistance. Firstly, we observe a systematic cascade of quantized states with different Chern numbers (including 0, -1, -2, and possibly -3; see Extended Data Fig. 3) at the same charge density, which is distinct from a fixed non-zero Chern number of AHC that is determined by the integrated Berry curvature over the occupied states. Secondly, the anomalous Hall signal is much smaller than $h/e^2$ in the mWC, indicating the dominant role of normal Hall signal in determining the quantized Hall resistance at non-zero magnetic field.

In conclusion, we observed evidence of Wigner and metallic Wigner crystals in rhombohedral graphene, enriching the emergent quantum phenomena in this highly tunable material platform, and calling for future investigations into the relationship between various correlated and topological ground states and their underlying mechanisms.